\documentclass{article}

\usepackage{arxiv}

\usepackage[utf8]{inputenc} 
\usepackage[T1]{fontenc}    
\usepackage{hyperref}       
\usepackage{url}            
\usepackage{booktabs}       
\usepackage{amsfonts}       
\usepackage{nicefrac}       
\usepackage{microtype}      
\usepackage{lipsum}
\usepackage{amsmath}
\usepackage[T2A,T1]{fontenc}
\usepackage[utf8]{inputenc}
\usepackage[russian,english]{babel}
\usepackage{cases}
\usepackage{graphicx}
\usepackage{wrapfig}
\usepackage[figurename=Fig.]{caption}

\title{Translation: About a Method of Obtaining Volumetric Images by Means of X-ray Radiation}

\begin{document}
\maketitle
\begin{center}
\foreignlanguage{russian}{Известия Киевского Ордена Ленина Политехнического Института}\\
\foreignlanguage{russian}{Том 22, 154 (1957)}\\

\hspace{1cm}

\foreignlanguage{russian}{О МЕТОДЕ ПОЛУЧЕНИЯ ОБЪЕМНЫХ ИЗОБРАЖЕНИЙ
ПРИ ПОМОЩИ РЕНТГЕНОВСКОГО ИЗЛУЧЕНИЯ
}\\
\foreignlanguage{russian}{С. И. Тетельбаум}\\
\hspace{1cm}
\hrule

Bulletin of the Kiev Polytechnic Institute

Volume 22, 154 (1957) 

ABOUT A METHOD OF OBTAINING VOLUMETRIC IMAGES BY MEANS OF X-RAY RADIATION

S.I. Tetelbaum

Translation: Alex Gustschin\footnote{Chair of Biomedical Physics, Department of Physics and Munich School of Bioengineering, Technical University 
of Munich, 85748, Garching, Germany, alex.gustschin@ph.tum.de}

\hspace{1cm}
\hrule
\end{center}

\section*{Introduction}
State-of-the art x-ray technology allows to obtain shadow images of an object under investigation. The shadow-like character of the image on the fluorescent screen or the recorded x-ray film results from intensities of radiation transmitting the object, which depend on the effective attenuation of all elements of the object along one straight line of every ray of the beam.

The observed radiation intensity distribution $f$ in an image in a certain plane in space depends on the attenuation coefficient $F(x,y,z, \lambda)$ of the object’s elements, which is a function of the geometrical coordinates and the wavelength of the radiation $\lambda$ as well as other experimental conditions.

A volumetric representation of the studied object can be obtained if the function $F$ (characterized by the local attenuation coefficient of every element of the three-dimensional object) is determined by experiment. There is some functional dependence between functions $f$ and $F$:

\begin{equation} \label{eq:1}
f = \Phi(F).
\end{equation}

In conventional methods of x-ray technology this relation does not provide a unique solution for the function $F$ and the observed shadow image can be --- generally speaking --- generated by some high plurality of different objects resulting in the same attenuation.

In this article --- based on some general ideas on possible compensations of image distortions generated by optical and analog instruments [1] --- a method is given to obtain three-dimensional images of objects with x-rays or other kind of sufficiently penetrating radiation. Likewise, an image of a sufficiently thin layer of the object which is not altered by features of apposed layers can be obtained.

For this purpose, the conditions of the experiment are chosen such that the observed intensity distribution (which can be measured directly or provided by previously taken radiographs or fluorographs) clearly defines the object, i.e. so that the equation of the kind (\ref{eq:1}) has a single solution. In this case, using the inverse operator

\begin{equation} \label{eq:2}
F = \Phi^{*}(f)
\end{equation}

it is possible to find the required function $F$ which determines the local attenuation coefficient in every element of the three-dimensional object.  

\section*{Integral Equation of the Problem}
\label{sec:headings}

\begin{figure} 
  \centering
  \includegraphics[width=10cm]{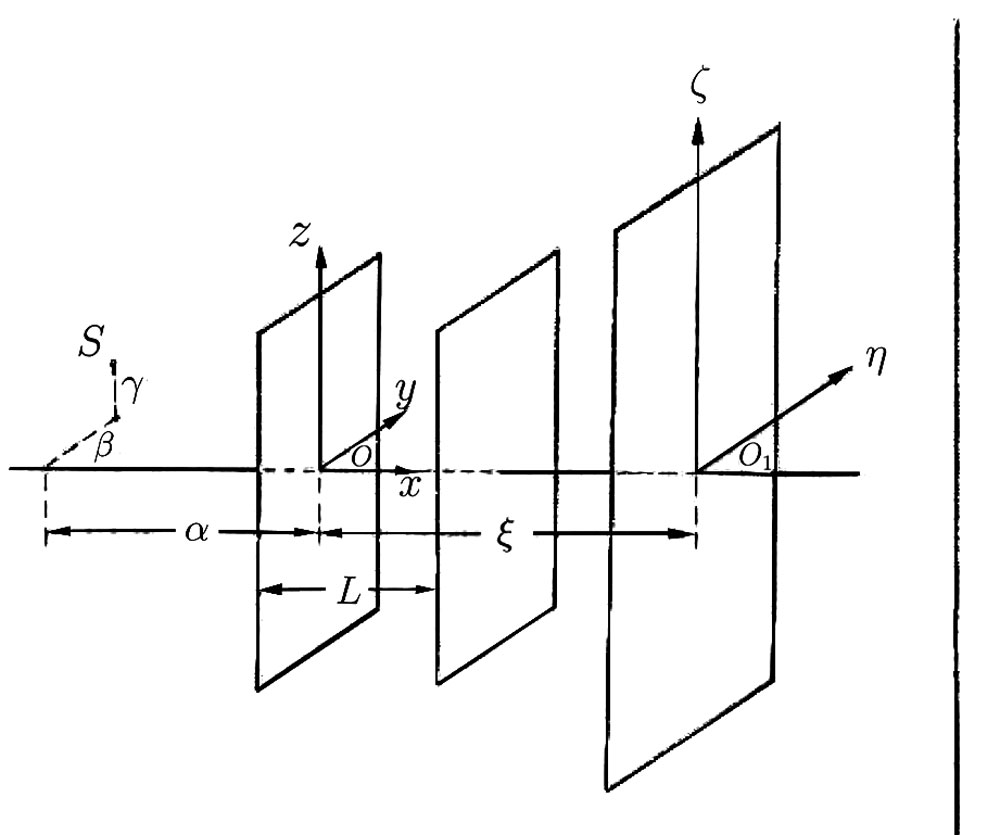}
  \caption{ }
  \label{fig:fig1}
\end{figure}

Let the point-like, non-directional source $S$ (Fig. \ref{fig:fig1}) with known coordinates $\alpha, \beta, \gamma$ emit x-rays or other penetrating radiation with spectral distribution $\phi_1(\lambda)$. The object---confined in three dimensions---is placed between the planes at $x=0$ and $x=L$. The intensity of the radiation is determined in a plane of observation $\eta O_1 \zeta$ parallel to the planes mentioned before.

We will characterize the object with the local attenuation coefficient $F(x,y,z, \lambda)$. Let’s redefine function $F$ in the space between $x=0$ and $x=L$ and set $F=0$ outside the object.
At this fixed geometry of the system the distribution of radiation intensity (up to a constant factor) will be:

\begin{equation} \label{eq:3}
F(\eta, \zeta, \lambda) = \frac{\phi_1(\lambda) \exp \Bigl\lbrack  -\varphi(\alpha, \beta, \gamma, \eta, \zeta, \xi) \int\limits_{0}^{L} F(x,y,z, \lambda) \ dx \Bigr\rbrack}{(\xi + \alpha)^2 [\varphi(\alpha, \beta, \gamma, \eta, \zeta, \xi)]^3}, 
\end{equation}

\begin{equation*}
\begin{aligned}
 \varphi(\alpha, \beta, \gamma, \eta, \zeta, \xi) = \sqrt{1+\Bigl(\frac{\eta-\beta}{\xi + \alpha}\Bigr)^2 + \Bigl(\frac{\zeta-\gamma}{\xi + \alpha}\Bigr)^2},
 \end{aligned}
\end{equation*}
where

\begin{equation*}
\begin{aligned}
y = \beta\Bigl(1+\frac{x+\alpha}{\xi - \alpha}\Bigl) + \eta \frac{x+\alpha}{\xi + \alpha};  
z = \gamma \Bigl(1+  \frac{x+\alpha}{\xi - \alpha}\Bigl) + \zeta \frac{x+\alpha}{\xi + \alpha}
 \end{aligned}
\end{equation*}

Equation (\ref{eq:3}) considers the attenuation of the radiation caused by absorption in the object, weakening due to the divergent cone of the beam as well as the inclined incidence of the rays onto the flat observation plane.
The integral equation (\ref{eq:3}) does not satisfy the criteria for a unique, continuous solution and function $F$ cannot be uniquely determined from the known function $f$. In conventional x-ray technology an additional spectral averaging occurs so that the brightness of the fluorescent screen or the darkening of the film during a static exposure is determined by the expression:

\begin{equation} \label{eq:4}
B(\eta, \zeta) = \Theta {\Bigl\lbrack \int\limits_{0}^{\infty} \phi_2 (\lambda) f(\eta, \zeta, \lambda) \ d\lambda \Bigr\rbrack} .
\end{equation}

In this expression $\phi_2 (\lambda)$ and $\Theta$ are describing the spectral sensitivity and the darkening function of the x-ray film or the respective characteristics of the screen.
In order to obtain the relations allowing to unambiguously determine the absorption coefficient in each element of the volume, we will determine the function $f$ with continuous change in spatial relation of the radiation source, object and the plane of detection within some boundaries. In this case the respective integral equation will satisfy the conditions for a unique continuous solution and the function of interest $F$ can be determined unambiguously. 
If, for example, any coordinate of the radiation source during the examination is altered we get instead of equation (\ref{eq:3}) an integral equation:

\begin{equation} \label{eq:5}
F(\eta, \zeta, \sigma, \lambda) = \frac{\phi_1(\lambda) \exp \Bigl\lbrack  -\varphi(\alpha, \beta, \gamma, \eta, \zeta, \xi) \int\limits_{0}^{L} F(x,y,z, \lambda) \ dx \Bigr\rbrack}{(\xi + \alpha)^2 [\varphi(\alpha, \beta, \gamma, \eta, \zeta, \xi)]^3}, 
\end{equation}

where $\sigma$ is equal $\alpha, \beta$ or $\gamma$, satisfying the conditions of a unique solution. A convenient method to change the relative arrangement might also be the rotation of the object around a stationary axis.

\section*{Case of object rotation}

\begin{figure}
  \centering
  \includegraphics[width=8cm]{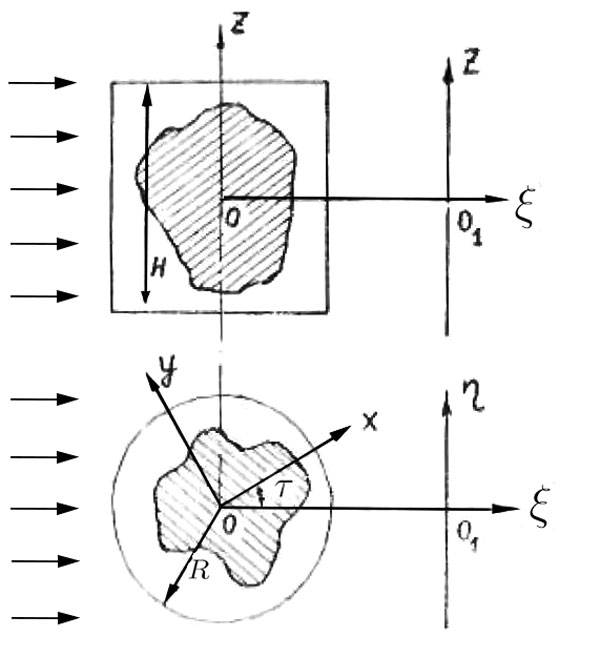}
  \caption{ }
  \label{fig:fig2}
\end{figure}

Let's consider a simplified case where the source is sufficiently distanced from the object. Let the object under study, whose cross sections (hatched) are shown in mutually perpendicular planes $\xi O_1 z$ and $xOy$) in Fig. \ref{fig:fig2}, be irradiated by a uniform, parallel beam of penetrating radiation propagating in the direction $O \zeta$. Let's determine the function $F(x,y,z,\lambda)$ in the cylinder of radius $R$ and height $H$ enclosing the object and setting $F=0$ outside the object in the cylinder. Let’s find the intensity distribution of the radiation $I$ in  the observation plane $\eta O_1 z$ during the rotation of the object around the axis $O z$ with angular velocity $d \tau / dt$. Up to a constant factor we get:

\begin{equation} \label{eq:6}
f(\eta, z, \tau, \lambda) = \phi_1(\lambda) \exp \Bigl\lbrack  - \int\limits_{-\sqrt{R^2-\eta^2}}^{\sqrt{R^2-\eta^2}} F(x,y,z, \lambda) \ d\xi \Bigr\rbrack, 
\end{equation}

where

\begin{equation*}
x = \xi \cos\tau + \eta \sin\tau; y = - \xi \sin \tau + \eta \cos\tau;
\end{equation*}

and $\tau$ is the angle between $O\xi$ and $Ox$ describing the rotation of the object. 

The integral equation (\ref{eq:6}) likewise satisfies the conditions of a unique, continuous solution. If, for example, the observation is conducted in a relatively narrow interval of wavelengths around $\lambda_1$ so that $F(x,y,z,\lambda_1) = F_1(x,y,z)$ and respectively $f(\eta, z, \tau, \lambda_1) = f_1(\eta, z, \tau)$ the solution of equation (\ref{eq:6}) becomes\footnote{The given calculation was performed by B. I. Korenblum}:

\begin{equation} \label{eq:7}
\begin{aligned}
 F(x,y, z) = - \frac{1}{4 \pi^2} \lim_{\delta \to + 0} \int  \limits_{0}^{2\pi} \int  \limits_{-\infty}^{+\infty}
 \frac{\lbrack r \sin{(\varphi +\tau)} - \eta \rbrack^2 - \delta^2 }{\lbrack r \sin{(\varphi +\tau)} - \eta \rbrack^2 +
 \delta^2} \ln{ f_1(\eta, z, \tau)} d\eta d\tau,\\\
  \end{aligned}
\end{equation}

where
\begin{equation*}
\begin{aligned}
 x = r \cos{\varphi}; y= r \sin{\varphi}; 0\leq r<R.
  \end{aligned}
\end{equation*}

The operation of finding function $F$ from the experimentally determined function $f$ can be executed automatically with the help of a respective analog computing device. For a hardware-based determination of function $F$ the equation (\ref{eq:6}) can be rearranged to

\begin{equation} \label{eq:8}
\begin{aligned}
 F_1(x,y, z) = \frac{1}{4 \pi^2} \int  \limits_{0}^{2\pi} d\tau \int  \limits_{-R}^{+R}
 \frac{\frac{\partial}{\partial\eta} f_1(\eta, z, \tau)}{\eta - r \sin{(\varphi+\tau)}} d\eta ,\\\
  \end{aligned}
\end{equation}

which is derived from (\ref{eq:7}) taking the limit. 
Using (\ref{eq:8}) it should be noted that the order of integration cannot be interchanged. Deriving (\ref{eq:8}) some natural boundary conditions connected to a hardware-based determination of function $F$ were taken into account (continuous derivatives up to the third). 

It should be noted that the considered condition of a parallel beam (which was done for simplicity) is not mandatory. In the case of a relatively close placement of the radiation source the integral equation corresponding to the rotating object has also a unique solution.

With a sufficiently fast operation of obtaining a volumetric “frame” an observation of a moving objects can be realized, in which $F$ also depends on time $t$.

It is worth noting another principal possibility to improve the differentiation of some parts of the object, distinguished by their different dependence of the local attenuation coefficient on the wavelength. When $F(x,y,z)$ is determined e.g. in three different, relatively narrow wavelength intervals and the respective images are retrieved, they can be colored with primary colors and merged. The result will be a conditionally colored image, which could provide, e.g. in medicine, additional diagnostic possibilities.

Since the function $f$ has to be acquired, it can be e.g. implemented with a fluorographic frame recorder taking a series of shadow images for discrete values of $\tau = \tau_0 + n\Delta\tau (n = 1,2,3,..., m)$.
It should be considered that reducing the interval $\Delta\tau$ and increasing $m$ the achievable detail reproduction of the function $F$ is increased. 

\begin{figure}
  \centering
  \includegraphics[width=8cm]{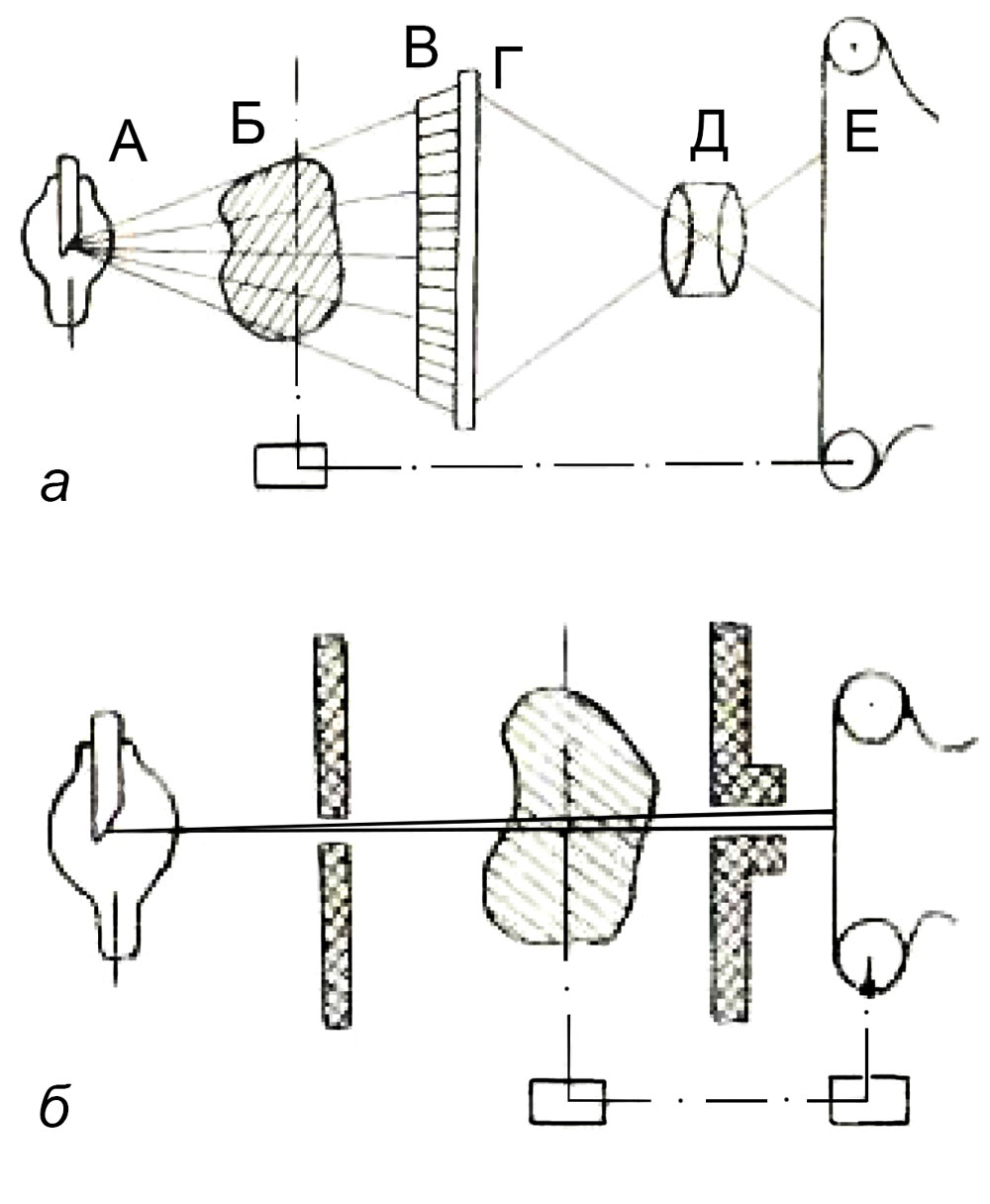}
  \caption{ }
  \label{fig:fig3}
\end{figure}

A schematic drawing of the setup is shown in Fig. \ref{fig:fig3}a. Here is: A---source of radiation, \foreignlanguage{russian}{Б}---object, B---grid-like lead filter absorbing the scattered rays, \foreignlanguage{russian}{Г}--- fluorescent screen (equipped with a light amplifier, if the total radiation dose should be decreased), \foreignlanguage{russian}{Д}--- objective, E--- film. 

The kinematic coupling between the object rotation drive and the film translation device is indicated by the dot-dashed lines. Studying a thin layer the recording of the function $f(\eta, \tau)$ can be realized in the form of an encoded radiograph on a uniformly moving film and a uniform object rotation. The respective setup is depicted in Fig. \ref{fig:fig3}b. The cross-sections of the lead slit collimators are cross-hatched.

\section*{Conclusions}
The article demonstrates a principle possibility of determining the local attenuation coefficient in every element of a three-dimensional object using penetrating radiation. The task is expressed as an integral equation and the experimental conditions are chosen such that this equation has a unique solution. The possibility of obtaining volumetric x-ray images and radiographs of thin layers is shown as well as the use of conditional color contrasting.
The considered method is of promising interest for x-ray and gamma-based defect detection as well as for medical diagnostics.

\bibliographystyle{unsrt}  

\begin{center}
    -END OF ORIGINAL ARTICLE-
\end{center} 

\section*{Translator's Note}
\label{sec:others}

The present translated article from 1957 appears remarkable as it develops the idea of x-ray computed tomography and discusses its possibilities even prior to the work of Cormack in 1963. It was hardly noticed in literature until today due to limited accessibility and the language barrier. The interested reader should be referred to another article\footnote{B. I. Korenblum, S. I. Tetelbaum, A. A. Tyutin, About one scheme of tomography, Bulletin of the Institutes of Higher Education - Radiophysics 1, 151-157 (1958)} by Korenblum, Tetelbaum and Tyutin from 1958 suggesting and deriving a scheme for fan-beam CT of a thin slice. A translation of the later is in progress and will be made available soon.

Since the original article does not have an abstract, the conclusion part with an introductory sentence was chosen as abstract for the online record. The figures have been improved (mainly the display of the variables) for a better recognition.

I would like to thank Fabio De Marco for bringing the work to my attention and further discussion and research on original sources. I am grateful to the staff members of the \textit{Scientific and Technical Library of Igor Sikorsky Kyiv Polytechnic Institute} for helping to find copies of the original work and the references.

\section*{Appendix A}

Scans of the original work in Russian language are provided bellow. 

\begin{figure}
  \centering
  \includegraphics[width=15cm]{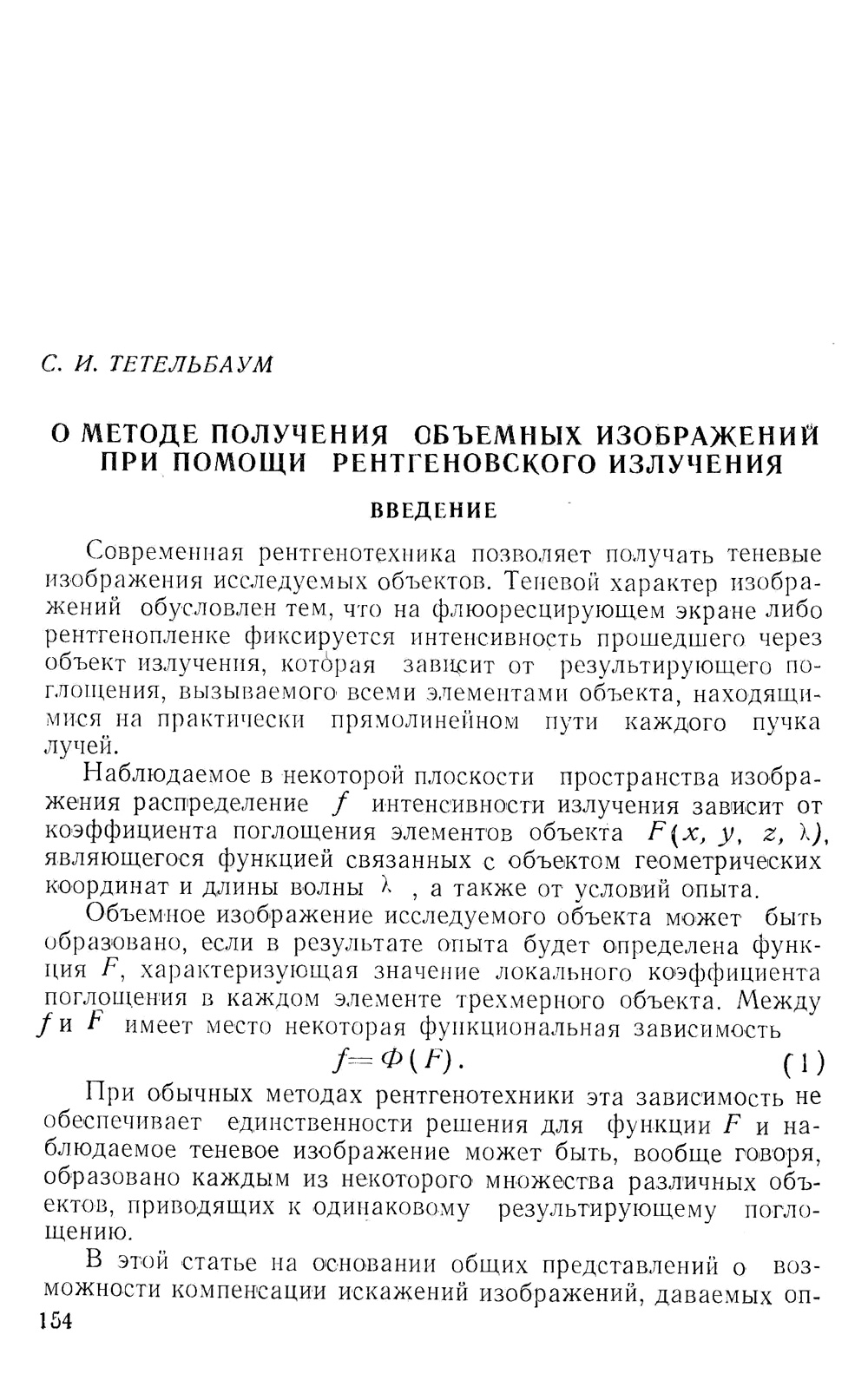}
\end{figure}

\begin{figure}
  \centering
  \includegraphics[width=15cm]{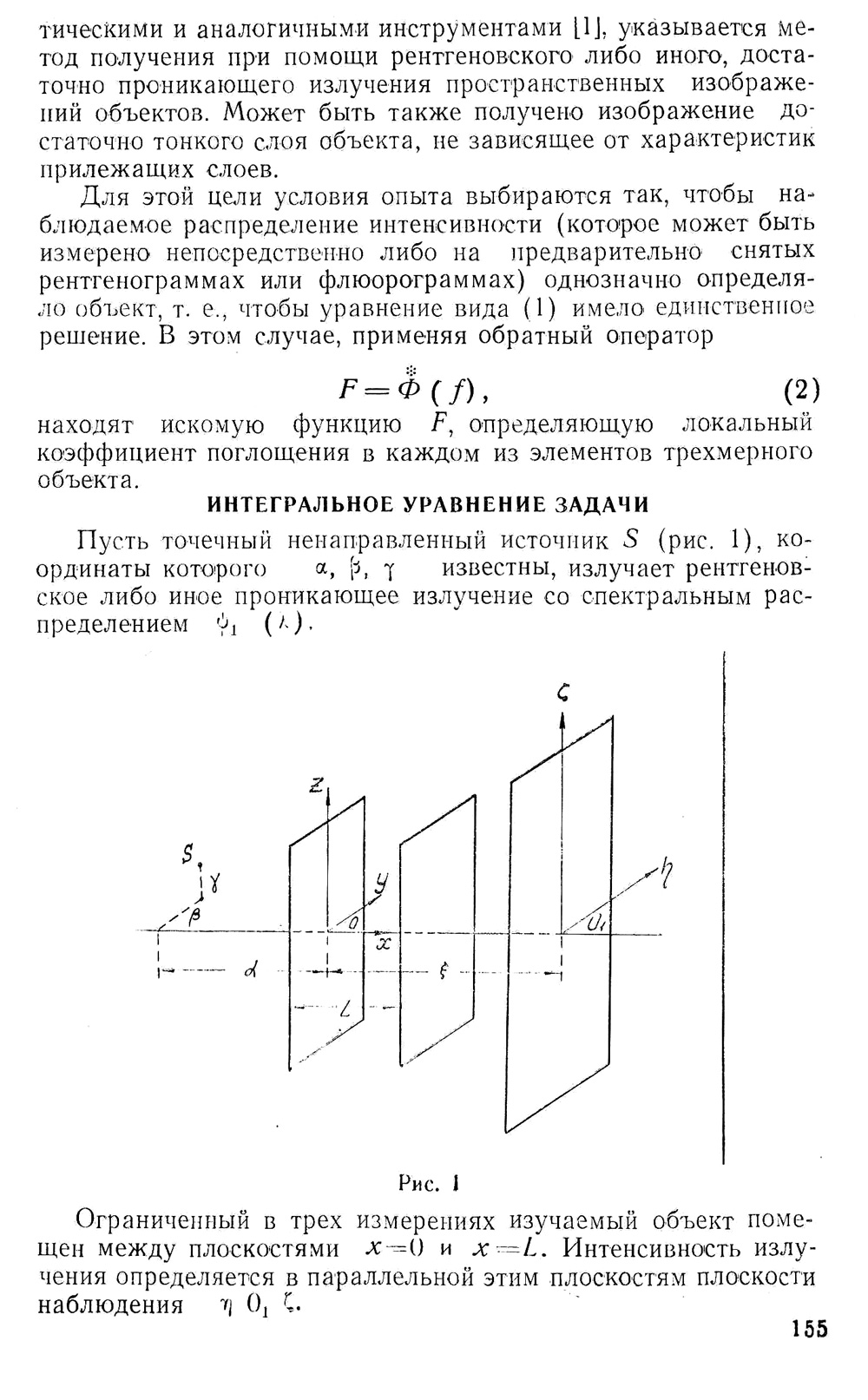}
\end{figure}

\begin{figure}
  \centering
  \includegraphics[width=15cm]{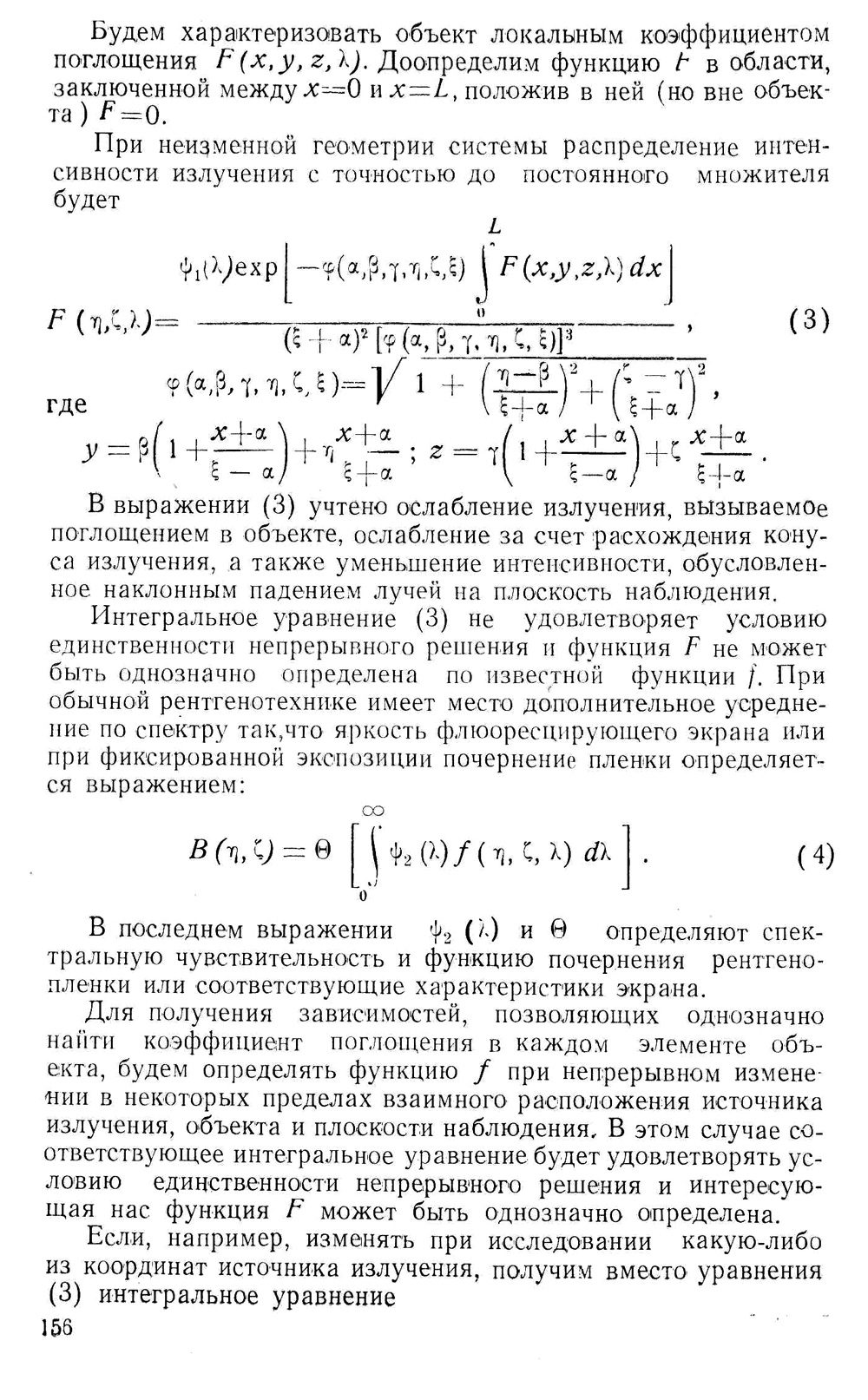}
\end{figure}

\begin{figure}
  \centering
  \includegraphics[width=15cm]{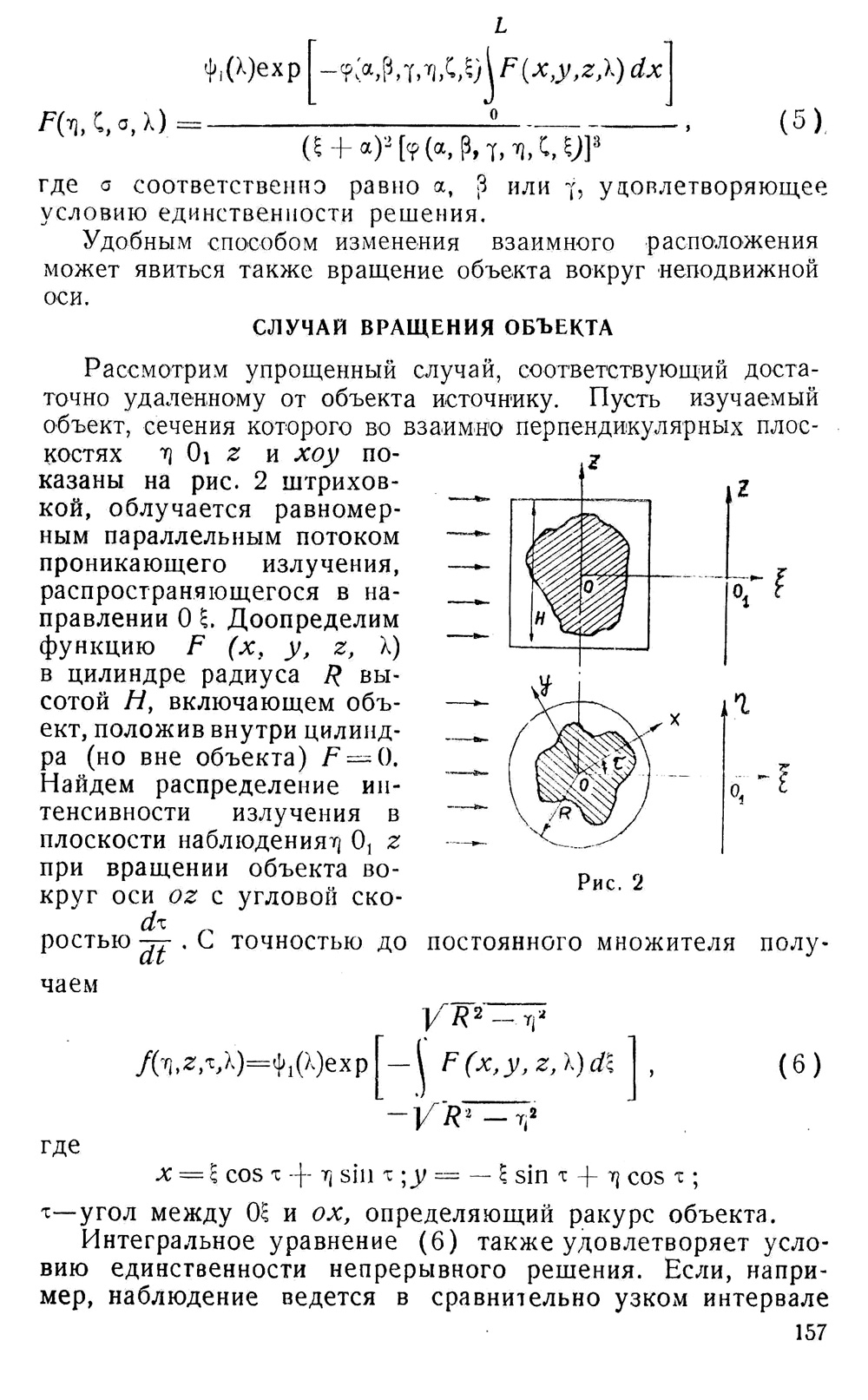}
\end{figure}

\begin{figure}
  \centering
  \includegraphics[width=15cm]{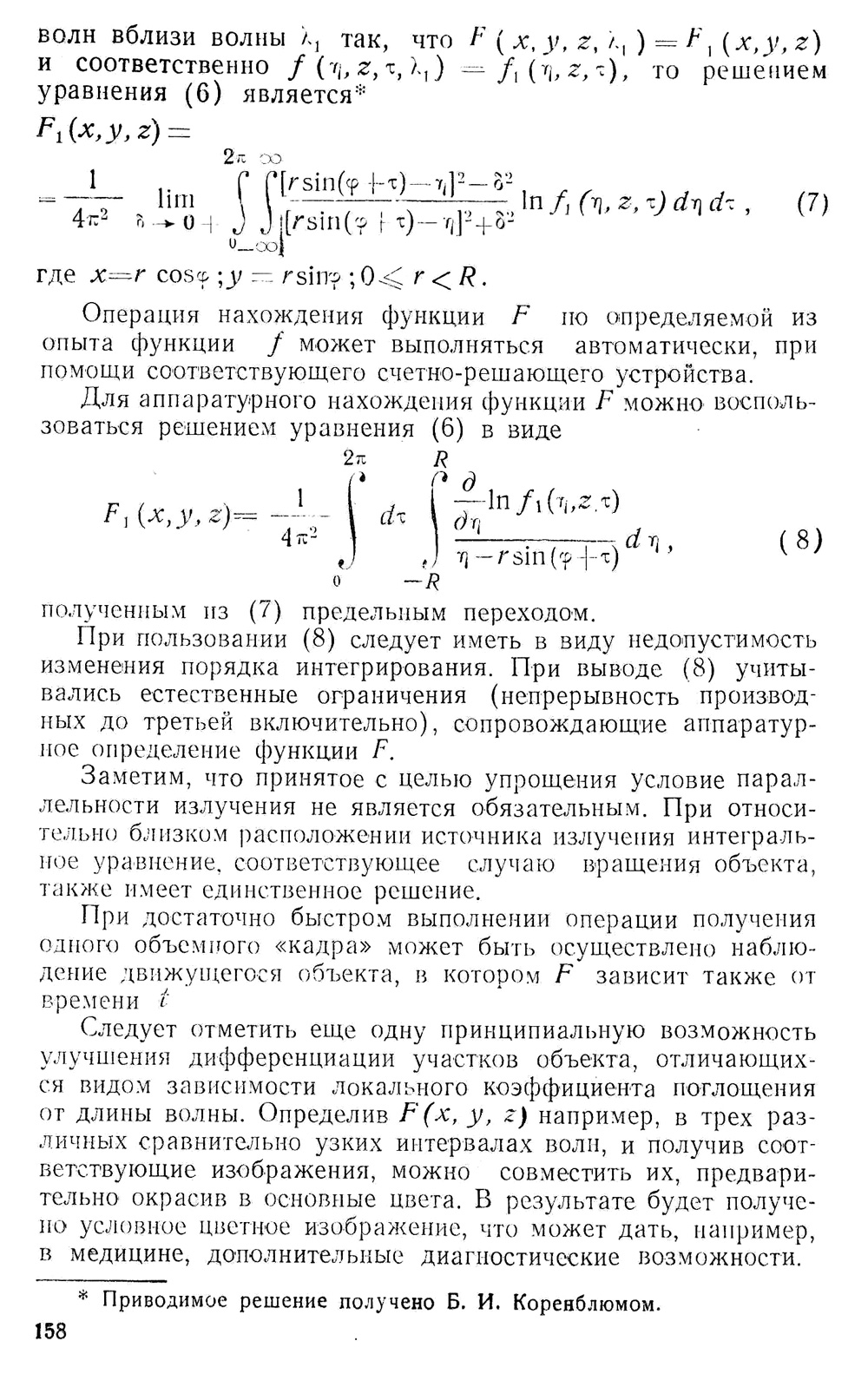}
\end{figure}

\begin{figure}
  \centering
  \includegraphics[width=15cm]{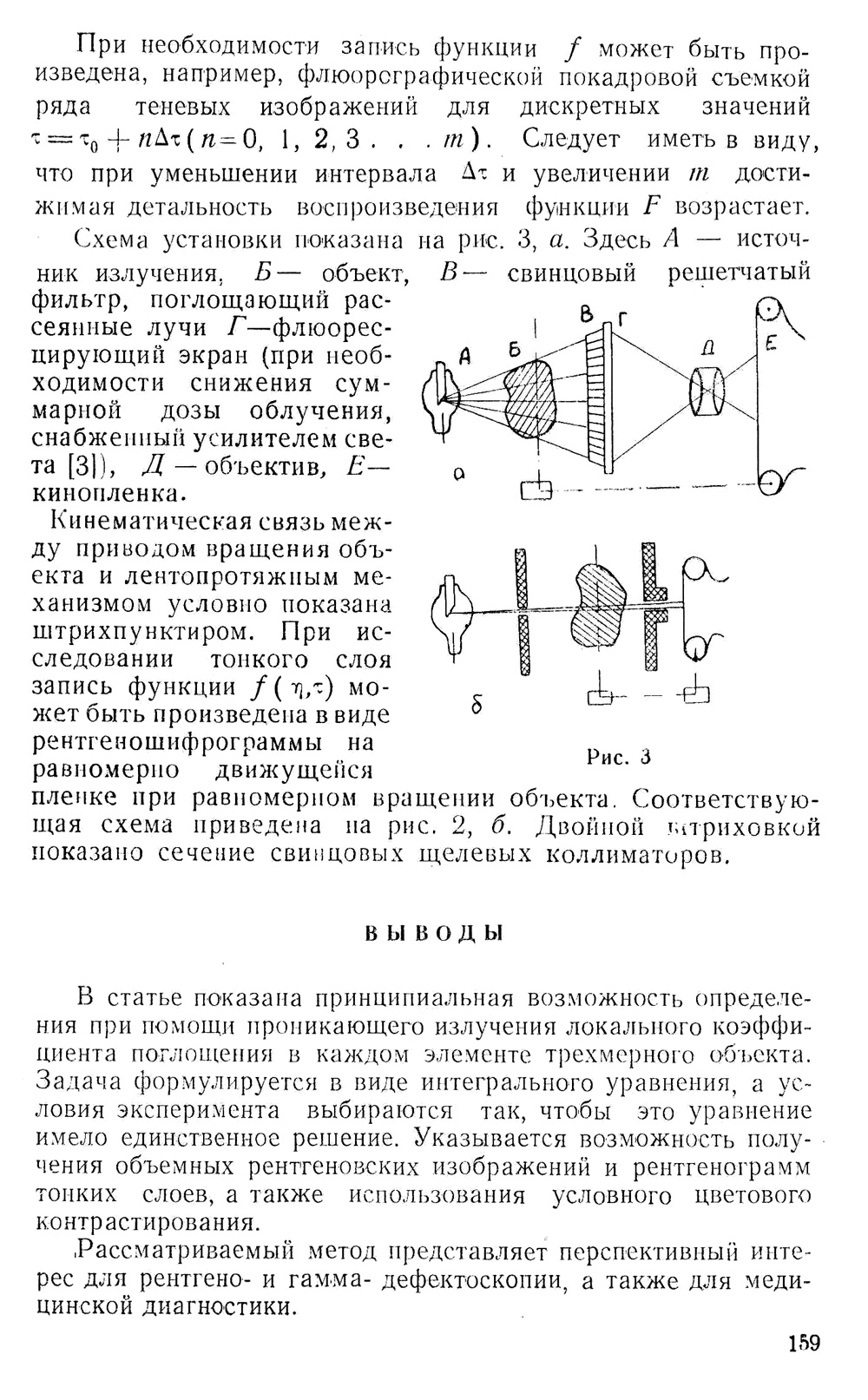}
\end{figure}

\begin{figure}
  \centering
  \includegraphics[width=15cm]{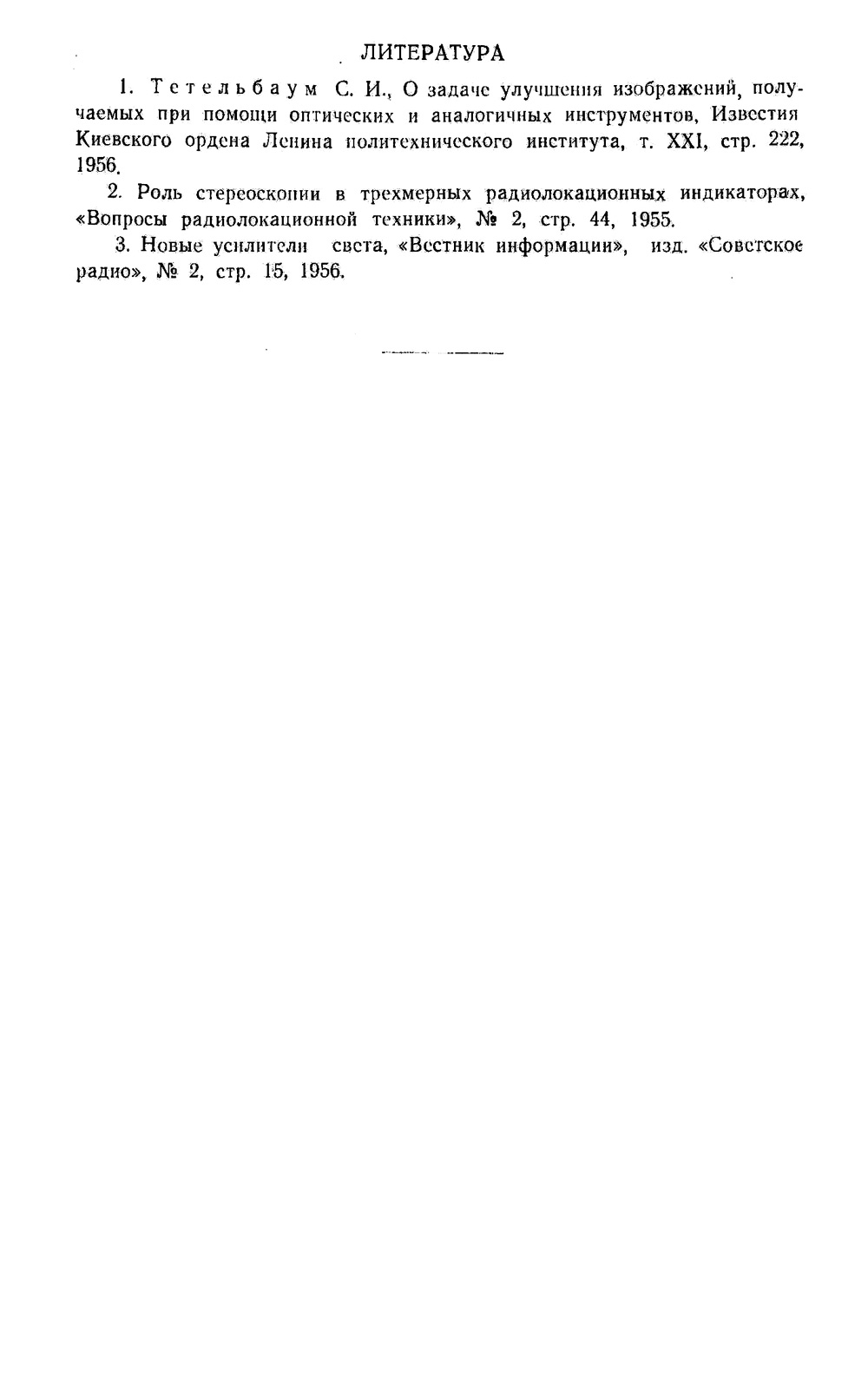}
\end{figure}

\end{document}